\documentclass[journal]{IEEEtran}

\usepackage{amsfonts,subfigure,multicol,color,verbatim,
graphicx,cite,epsfig,amssymb,amsmath,cases,bm,algorithm,
algorithmic,xcolor,multirow,url,stfloats,enumerate,dsfont,bbm,tabularx}

\newtheorem{theorem}{Theorem}
\newtheorem{lemma}[theorem]{Lemma}
\newtheorem{proposition}[theorem]{Proposition}

\newtheorem{Definition}{Definition}

\begin{document}
\title{Mode Selection and Resource Allocation in Sliced Fog Radio Access Networks: A
Reinforcement Learning Approach}

\author{Hongyu~Xiang, Mugen~Peng,~\IEEEmembership{Fellow,~IEEE}, Yaohua Sun, and Shi Yan,~\IEEEmembership{Member,~IEEE}
\thanks{
Copyright (c) 2015 IEEE. Personal use of this material is permitted. However, permission to use this material for any other purposes must be obtained from the IEEE by sending a request to pubs-permissions@ieee.org.

Hongyu~Xiang (e-mail: xhyou@bupt.edu.cn),  Mugen~Peng (e-mail: pmg@bupt.edu.cn), Yaohua~Sun (e-mail: sunyaohua@bupt.edu.cn), and Shi~Yan (e-mail: yanshi01@bupt.edu.cn) are with the State Key Laboratory of Networking and Switching Technology, Beijing University of Posts and Telecommunications, Beijing, China.}
}

\maketitle

\begin{abstract}
The mode selection and resource allocation in fog radio access networks (F-RANs) have been advocated as key techniques to improve spectral and energy efficiency. In this paper, we investigate the joint optimization of mode selection and resource allocation in uplink F-RANs, where both of the traditional user equipments (UEs) and fog UEs are served by constructed network slice instances. The concerned optimization is formulated as a mixed-integer programming problem, and both the orthogonal and multiplexed subchannel allocation strategies are proposed to guarantee the slice isolation. Motivated by the development of machine learning, two reinforcement learning based algorithms are developed to solve the original high complexity problem under traditional and fog UEs' specific performance requirements. The basic idea of the proposals is to generate a good mode selection policy according to the immediate reward fed back by an environment. Simulation results validate the benefits of our proposed algorithms and show that a tradeoff between system power consumption and queue delay can be achieved.
\end{abstract}


\IEEEpeerreviewmaketitle

\section{Introduction}

To handle diverse use cases and business models, a new technology called network slicing
has been investigated extensively for fifth generation (5G)\textcolor[rgb]{1.00,0.00,0.00}{\cite{NS}}.
In the concept of network slicing,
the network slice instances are orchestrated and chained
by a set of network functions to provide customized services.
By enabling flexible support of various applications,
network slicing benefits 5G networks in a cost-efficient way.
As an important part of network
slicing, network slicing in radio access networks (RANs) has been studied to further improve end-to-end
network performance\textcolor[rgb]{1.00,0.00,0.00}{\cite{RANS5}}.

Although network slicing is a good solution to meet service requirements in 5G,
there are remarkable challenges to be solved.
Traditional core network slicing methods are business-driven only,
which neglect characteristics of the RAN.
However, network slicing in different network architectures
are different, like in heterogeneous networks or
cloud RANs (C-RANs)\textcolor[rgb]{1.00,0.00,0.00}{\cite{RANS2,Tony2}}.
Jointly considering characteristics of RANs and network slicing can be beneficial.
Second, the performance requirements of emerging applications become
more stringent. To achieve huge capacity, massive connections and ultra-low latency,
resource allocation should be elaborately designed, which includes not only
radio but also caching and computing resources.
Third, as indicated in TS 38.300\textcolor[rgb]{1.00,0.00,0.00}{\cite{TS38300}},
it should be possible for a single RAN node to support multiple slices.
Due to the differentiated capability of each node, the node association strategy
in a sliced RAN becomes critical.

Meanwhile, fog-RANs (F-RANs) have been considered as a revolutionary paradigm
to tackle performance requirements in 5G\textcolor[rgb]{1.00,0.00,0.00}{\cite{FRAN}}. By exploiting the edge caching and computing,
capacity burdens on fronthaul are alleviated
and end-to-end latency is shortened.
According to desired performance, each
user equipment (UE) in a F-RAN can select a proper
communication mode,
which includes C-RAN mode, fog-radio access point (F-AP) mode, device-to-device (D2D) mode.
With adaptive mode selection and interference
suppression,
services and applications, such as the
industrial Internet, health monitoring and Internet
of vehicles, can be well supported.

To exploit the prospect of network slicing in F-RANs,
a hierarchical RAN slicing architecture is presented in this paper.
The proposed architecture shown in Fig. \ref{sys}
takes full advantages of both F-RANs and network slicing.
According to the decomposition principle of the control and
data planes, the high power node (HPN) in the network access layer
executes the functions of the control plane, including
control signaling and system
broadcasting information delivery for accessed traditional UEs and fog-UEs (F-UEs).
With radio resource control connections established, the network slice selection assistance information\textcolor[rgb]{1.00,0.00,0.00}{\cite{TS38300}} is utilized to help traditional UEs and F-UEs
for network slice selection.
Numerous modes are provided in the data plane for a differentiated handling of traffic.
Specially, the remote radio heads (RRHs) are cooperated with each other in the baseband unit (BBU) pool,
which provides the C-RAN mode in the data plane.
Thanks to the fog computing, F-APs are used to process local
collaboration radio signal and D2D mode can be
further triggered to meet performance requirements.

In the RAN slicing architecture, mode selection and resource allocation are critical for improving
performance of network slices. To achieve a high data rate, UEs should associate with
RRHs to leverage large scale centralized signal processing in the BBU pool.
To alleviate transmission burdens on fronthaul and save system power,
local data processing should be available, which are enabled by
F-APs and F-UEs.
Consequently, for UEs with different performance requirements,
advanced mode selection are in need.
Note that the data transmission under different modes
would consume not only radio but also computing resource.
Like in C-RAN mode, centralized processing and large-scale
collaborative transmission requires global
coordination, scheduling and control, of which the computing
complexity typically increases polynomially with the network
size\textcolor[rgb]{1.00,0.00,0.00}{\cite{FRAN}}. Hence it is important to coordinate the computing and radio resource.
To meet the performance requirements of traditional UEs and F-UEs,
both multi-dimensional resource management
and communication mode selection in sliced F-RANs should
be tackled elaborately.
Considering their coupling, a joint
optimization of mode selection and
resource allocation is essential.
To determine the best mode selection and coordinate the multi-dimensional resource,
intelligent
decision-making mechanisms are promising, which consider
the channel states of different modes, the computing load at each F-AP, the performance requirements
of traditional UEs and F-UEs and the total power consumption.

Based on the aforementioned characteristics of mode selection and resource allocation,
the joint optimization solution to system power minimization
in sliced F-RANs is researched in this paper.

\subsection{Related Work}

F-RANs have emerged as a promising
5G RAN that can satisfy diverse quality of
service (QoS) requirements in 5G.
With coordination among the communication,
computation and caching, QoS requirements like high spectral efficiency,
high energy efficiency and low latency for different service types can be met.
Many studies on F-RANs have been
conducted, like computation offloading in\textcolor[rgb]{1.00,0.00,0.00}{\cite{FR0}},
and edge caching strategies in\textcolor[rgb]{1.00,0.00,0.00}{\cite{FR1,FR2}}.
In\textcolor[rgb]{1.00,0.00,0.00}{\cite{FR0}}, the impact of fog computing on energy consumption and delay performance are investigated. With queuing models established, a multi-objective optimization problem considering energy consumption, execution delay and payment cost is formulated.
Using the scalarization method
and interior point method, superior performance over the existing schemes
is achieved. In\textcolor[rgb]{1.00,0.00,0.00}{\cite{FR1}}, a joint optimization of caching and user association is studied. By decomposing the original problem, a distributed algorithm based on the
Hungarian method is proposed. Simulation results show that with an efficient caching policy,
the average download delay can be significantly reduced. In\textcolor[rgb]{1.00,0.00,0.00}{\cite{FR2}},
a new metric called economical energy efficiency is adopted.
With cache status and fronthaul capacity considered, a resource allocation problem is formulated and solved by using fractional programming. Advantages of the proposed algorithm including system greenness improvement are confirmed.

There have also been numerous works on RAN slicing
that demands efficient
resource allocation, resource isolation and sharing\textcolor[rgb]{1.00,0.00,0.00}{\cite{SurvyeNS}}.
In\textcolor[rgb]{1.00,0.00,0.00}{\cite{RANS5}},
the application of network slicing in an ultra-dense RAN is studied.
To improve the quality of computation experience for mobile devices,
the design of computation offloading policies is investigated.
Considering the time-varying communication qualities and computation
resources, a stochastic computation offloading problem is formulated and then a deep reinforcement learning (DRL) framework is proposed,
which achieves a significant improvement in computation offloading performance compared with baseline policies.
In\textcolor[rgb]{1.00,0.00,0.00}{\cite{RANS2}}, a dynamic radio resource slicing framework
is presented for a two-tier heterogeneous wireless network.
By partitioning radio spectrum resources into different bandwidth slices for sharing, the framework achieves differentiated QoS
provisioning for services in the presence of network load dynamics.
In\textcolor[rgb]{1.00,0.00,0.00}{\cite{Tony2}},
two typical 5G services in a C-RAN are considered
and specific slice instances are orchestrated.
To maximize the cloud RAN operator's revenue, efficient approaches including successive convex approximation and semidefinite relaxation
are exploited. With acceptable time complexities, the proposed algorithm significantly saves system power consumption.
In\textcolor[rgb]{1.00,0.00,0.00}{\cite{SYH}},
hierarchical radio resource allocation is studied for RAN slicing in F-RANs,
where a global radio resource manager
performs a centralized subchannel
allocation while local radio resource managers allocate assigned resources to UEs
to facilitate slice customization.
In\textcolor[rgb]{1.00,0.00,0.00}{\cite{SPRM}},
the network slicing in multi-cell virtualized
wireless networks is considered. To maximize
the network sum rate, a joint BS assignment, sub-carrier, and power allocation algorithm
is developed. Simulation results demonstrate that
under the minimum required rate constraint of each slice,
the proposed iterative
algorithm outperforms the traditional approach, especially in the respect of the coverage improvement
and spectrum efficiency enhancement.
In\textcolor[rgb]{1.00,0.00,0.00}{\cite{NVHDT}},
the combinatorial optimization of multi-dimensional
resources in network slicing is investigated.
To deal with the dilemma between network provider
and tenants, a real-time resource slicing framework based on semi-Markov decision process
is developed, which considers the long-term return of the network provider
and the uncertainty of resource demands
from tenants. Taking advantages of deep dueling neural network,
the proposed framework can improve the performance of the system
significantly.
In\textcolor[rgb]{1.00,0.00,0.00}{\cite{JJRG}},
a novel spectral efficiency approach is proposed
to the allocation of resource blocks for different services.
By learning in advance whether resources is adequate
to provide service, unsuccessful allocation process is avoided.
Simulations show that the approach significantly improves the spectral efficiency
with respect to a single-slot based model.

Note that there still exist challenges in RAN slicing.
For example, the ever-increasingly complicated configuration
issues and blossoming new performance
requirements would be challenging in 5G,
since only predefined problems
can be dealt with by the network.
To realize an intelligent implementation of network slicing,
artificial intelligence has attracted particular attentions.
By enabling networks be capable of
interacting with environments,
a network can automatically
recognize a new type of application, infer an
appropriate provisioning mechanism and establish
a required network slice\textcolor[rgb]{1.00,0.00,0.00}{\cite{HZhang}}.
Meanwhile, with network scenarios becoming
heterogeneous and complicated,
cost-efficient and low-complexity algorithms
based on machine learning can be developed
for practical implementations\textcolor[rgb]{1.00,0.00,0.00}{\cite{AI6}}.
With network patterns and user behaviors learned and predicted,
an intelligent decision making system can be
established to improve the network performance.

There have been numerous works about applications of artificial intelligence and machine learning in wireless networks\textcolor[rgb]{1.00,0.00,0.00}{\cite{AI4}}. In\textcolor[rgb]{1.00,0.00,0.00}{\cite{AI1}}, the resource allocation
schemes for vehicle-to-vehicle (V2V) communications are investigated.
To avoid the large transmission overhead in the traditional centralized method,
a novel decentralized resource allocation mechanism
based on deep reinforcement learning is proposed. Each
V2V link or a vehicle acts as an independent agent and finds the optimal
sub-band and transmission power autonomously. Simulation results
showed that each agent can effectively learn to satisfy the
stringent latency constraints on V2V links while minimizing the interference
to vehicle-to-infrastructure communications. In\textcolor[rgb]{1.00,0.00,0.00}{\cite{AI2}},
applications of machine
learning to improve heterogeneous network traffic control are researched.
Based on traffic patterns at the edge
routers, a supervised deep learning system is trained.
Compared with benchmark routing strategy, the proposed system
outperforms in terms of signaling overhead, throughput,
and delay. In\textcolor[rgb]{1.00,0.00,0.00}{\cite{AI3}},
a DRL assisted resource allocation method
is designed for ultra dense networks.
The original multi-objective problem
is decoupled into two parts based on the general theory of DRL.
The spectrum efficiency (SE) maximization is utilized to build the deep neural network.
The residual objectives like energy efficiency (EE) and fairness,
are considered as the rewards to train
the deep neural network.
Simulation results show that, the proposed method significantly
outperforms the existing resource allocation algorithms in
term of the tradeoff among the SE, EE and fairness.
In\textcolor[rgb]{1.00,0.00,0.00}{\cite{AI5}},
the joint SE and EE optimization in cognitive radio networks
are studied and a deep-learning inspired message passing algorithm
is proposed. To learn the optimal parameters
of the algorithm, a feed-forward neural network is devised and an analogous
back propagation algorithm is developed. The simulation results show that the proposed
algorithm achieves a lower power consumption for secondary users accessing the licensed spectrum
while preserving the capacity of the primary users.

In this paper, we focus on the mode selection and resource allocation
in a sliced F-RAN, which is formulated as a
mixed integer programming.
To deal with the NP-hard problem, RL is adopted to
generate an efficient solution.
Combining the strength of both supervised and unsupervised
learning methods, the RL techniques have been
widely used in wireless networks\textcolor[rgb]{1.00,0.00,0.00}{\cite{SurvyeSYH}}.
In\textcolor[rgb]{1.00,0.00,0.00}{\cite{RL1}},
mode selection and resource allocation
in D2D enabled C-RANs are investigated
and a distributed approach based on RL is proposed, where D2D pairs perform self-optimization
without global channel state information.
In\textcolor[rgb]{1.00,0.00,0.00}{\cite{RL3}},
a decentralized and self-organizing mechanism based on RL techniques is
introduced to reduce inter-tier interference and improve spectral efficiency. Simulation results show that the proposed mechanism possesses better convergence properties and incurs less overhead than existing techniques.
To offload the traffic in a stochastic heterogeneous cellular network,
an online RL framework is presented in\textcolor[rgb]{1.00,0.00,0.00}{\cite{RL4}}.
By modeling as a discrete-time Markov decision process, the energy-aware traffic offloading problem is solved by a centralized Q-learning
algorithm with a compact state representation.

\subsection{Main Contributions}

Motivated by the benefits of machine learning, the uplink of a sliced F-RAN is concerned
in this paper. In particular, an optimization framework for
RAN slicing is presented, which takes
the queue stabilities of traditional UEs
and bit rate requirements of F-UEs into consideration.
Both orthogonal and multiplexed subchannel strategies are considered.
The main contributions of the paper are:

\begin{enumerate}
\item The joint optimization on mode selection and resource allocation in the uplink sliced F-RAN are investigated, where traditional UEs and F-UEs are served by constructed network slice instances. Both the orthogonal and multiplexed subchannel strategies are presented. Under different UEs' demands and limited computing resources, a system power minimization problem is formulated, which is stochastic and mixed-integer programming. Using the general Lyapunov optimization framework, this nonconvex optimization problem is transformed into a minimization of the drift-plus-penalty function, which can be further reformulated as a deterministic mode selection and resource allocation problem at each slot.

\item RL-based approaches are proposed to solve the reformulated mode selection and resource allocation problem. Unlike previous work in\textcolor[rgb]{1.00,0.00,0.00}{\cite{RANS2,Tony2,SYH}}, this paper applies the RL techniques to solve the drift-plus-penalty minimization under different subchannel allocation strategies. Specifically, communication modes are selected based on learned policies. Afterwards, transmission power of traditional UEs and F-UEs are derived by a generalized weighted minimum mean-square error (WMMSE) approach. Through the RL-based approaches, a long-term system performance optimization can be achieved.

\item The proposed approaches are evaluated under different conditions.
Impacts of different parameters like computing resource are evaluated.
By simulation, it is observed that the RL-based approach can provide real-optimal performance. By changing the value of the defined tradeoff parameter, tradeoff between traditional UEs' queuing delay and system power consumption can be controlled in a flexible and efficient way.

\end{enumerate}

The remainder of this paper is organized
as follows. Section II introduces the system model including the communication model
and computing model.
In Section III, the system power minimization problem is formulated
and transformed into a deterministic problem
based on the general Lyapunov optimization
framework. In Section IV, both the
orthogonal and multiplexed subchannel strategies
are considered, which enable different levels of slice isolation.
Corresponding RL-based algorithms are designed
to solve the deterministic problem.
Section V evaluates the
performance of the proposed algorithms, followed by the conclusions
in Section VI.

\section{System model}

The system model is elaborated in this section, including
the considered F-RAN model, communication model and computing model.

\subsection{The F-RAN model}
\begin{figure}[!htb]
\centering
\includegraphics[width=0.45\textwidth]{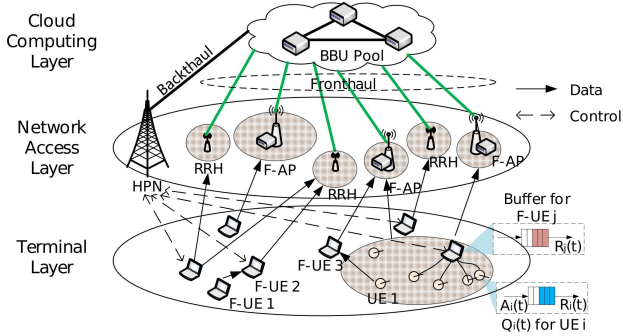}
\caption{The single antenna system model of the RAN slicing architecture, wherein network slices for traditional UEs and F-UEs are constructed.}
\label{sys}
\end{figure}

The scenario considered in this paper
is illustrated in Fig. \ref{sys}.
It assumes an F-RAN architecture
consisting of a terminal layer, a network access layer and a cloud computing
layer. In the cloud computing
layer, the BBU pool provides centralized signal
processing. And in the network access layer, there are
$L_1$ distributed RRHs connected with the BBU pool,
each of which is single-antenna.
There are also $M_0$ F-APs configured with $L_0 (L_0<L_1)$ antennas.
Owing to fog computing,
collaborative radio signal processing can not only be
executed in the centralized BBU pool but also
at distributed F-APs.
We also assume that the network operates in slotted time with
time dimension partitioned into decision slots indexed by $t \in \{0, 1, 2,...\}$

There are $K_0$ single-antenna traditional UEs
and $K_1$ single-antenna F-UEs in the terminal layer, whose sets
are denoted as $\mathcal{K}_0$ and $\mathcal{K}_1$, respectively.
Examples of traditional UEs include agricultural field monitoring sensors, and
industrial monitoring devices,
which desire low power consumption and have random bursty traffic arrivals.
F-UEs can be smartphones or laptops\textcolor[rgb]{1.00,0.00,0.00}{\cite{FRAN}},
which are always equipped with a large buffer.
To provide a high data rate for each F-UE,
a network slice instance is constructed, which is composed of
multiple modes and corresponding physical resource.
In the C-RAN mode, RRHs are cooperated for uplink data reception
and the BBU pool provides centralized signal detection and baseband processing.
Moreover,
F-APs are deployed for a local service to alleviate the burden on the fronthaul.
Similarly, both C-RAN mode and F-AP mode are available in the network slice instance specific for traditional UEs.
However, the objective is to maintain a low power consumption and stable transmission delay for traditional UEs.
In addition, F-UEs can benefit both network slice instances via the D2D mode.
Specially, F-UEs relay the data traffic of other F-UEs, which extends the coverage of the slice instance
for F-UEs; while in the slice instance for traditional UEs, F-UEs aggregate the data to allow more traditional UEs to be connected simultaneously.

There are $N$ subchannels to
be allocated, each of which is with bandwidth $W_0$.
In this paper, we consider
both the orthogonal and multiplexed subchannel strategies.
In the former, subchannel $n$ is allocated to at most one traditional UE $i$ or F-UE $j$,
which enables hard isolation between slice instances.
While in the latter, subchannel $n$ can be shared among multiple traditional UEs and F-UEs.
In this strategy, the isolation between the slice instances would be guaranteed with a sophisticated mode selection and resource allocation.
Although slice isolation
in current works is guaranteed mainly
through an orthogonal subchannel allocation strategy.
To achieve
higher spectrum utilization, it is still necessary to investigate
a multiplexed subchannel allocation strategy.

\subsection{The communication model}

To achieve the rate requirement $R_{th}$, F-UE $j$ should
connect to the proper F-AP/RRHs.
Denote the communication mode selection of F-UE $j$ at slot $t$
as $s_{j,m,n}^{TX}(t)$, which equals to $1$ when F-AP $m$ ($m \in \{1,2,...,M_0\}$) is selected and subchannel $n$ is allocated and equals to $0$ otherwise.
For notation simplicity, we define that $s_{j,0,n}^{TX}(t) =1$ in the case that C-RAN mode is selected (i.e., all RRHs are connected) and subchannel $n$ is allocated.
Suppose that the optimal linear detection, i.e., MMSE detection, is
employed, the uplink rate of F-UE $j$ at slot $t$ when $s_{j,m,n}^{TX}(t)=1$ is
\begin{equation}
\label{RateDef}
\begin{split}
&R_{j,m,n}(t)\!=\! W_0s_{j,m,n}^{TX}\!(t)\!\log(1\!+\!\frac{P_{j,n}(t)\|\textbf{v}_{j,m,n}^H(t)\textbf{h}_{j,m,n}(t)\|^2}{Int_{j,m,n}+\sigma ^2\|\textbf{v}_{j,m,n}(t)\|^2}),\\
&Int_{j,m,n} = \sum\limits_{k\neq j, k \in \mathcal{K}_0 \cup \mathcal{K}_1 } P_{k,n}(t) \|\textbf{v}_{j,m,n}^H(t)\textbf{h}_{k,m,n}(t)\|^2,
\end{split}
\end{equation}
where $P_{j,n}(t)$ is the transmission power of F-UE $j$ on subchannel $n$,
$\textbf{h}_{k,m,n}(t)$ is the channel vector
between UE $k$ and the F-AP $m$ on subchannel $n$, $\textbf{v}_{j,m,n}(t)$ is
the MMSE detection vector, and $\sigma ^2$ is the noise power.
Note that these channel vector data account for the antenna
gain, path loss, shadow fading, and fast fading together.

Similarly, the rate $R_{i}(t)$ of traditional UE $i$ can be obtained, $R_{i}(t) =\sum\limits_{m = 0}^{M_0+K_1} \sum\limits_{n=1 }^N R_{i,m,n}(t)$.
Besides guaranteeing a precise rate threshold $R_i^{min}$,
a stable queue backlog is also considered for traditional UE $i$ given its random traffic
arrival characteristics.
Let $Q_{i}(t)$ represent the queue backlog
for traditional UE $i$ in slot $t$. As shown in Fig. \ref{sys}, we have
the following expression for the dynamics of queue backlog $Q_{i}(t)$,
\begin{equation}
\label{eq:I3}
\begin{split}
Q_{i}(t+1)=  \max \{Q_{i}(t)-R_{i}(t),0\} + A_i(t),
\end{split}
\end{equation}
where $A_i(t)$ is the number of bits for traditional UE $i$
to be uploaded in time slot $t$.
Note that $A_i(t)$ varies over time and we have $\mathbb{E}\{A_i\} = \lambda_i $.
To minimize
the average queue backlog and maintain stability, we
seek to perform a queue-aware resource allocation. A
definition on the queue stability which bounds the
average queue backlog is described in (\ref{eq:C}).

\begin{Definition}
\label{stable}
\emph{(Queue stability\textcolor[rgb]{1.00,0.00,0.00}{\cite{II:R1}})}. The queue backlog $Q_{i}(t)$
which is a discrete time process would be mean-rate stable if
\begin{equation}
\label{eq:C}
\begin{split}
C0: \lim_{t \rightarrow \infty} \frac{\mathbb{E}\{|Q_{i}(t)|\}}{t} =0 , i \in \mathcal{K}_0
\end{split}
\end{equation}
\end{Definition}

Besides RRHs and F-APs, F-UE $j \in \mathcal{K}_1$ can be also selected
as serving nodes of UE $k$ ($s_{k,j,n}^{TX}(t)=1$).
Taking advantage of a large buffer,
an F-UE can help upload the data of other F-UEs and traditional UEs.
For example, F-UE $1$ in Fig. \ref{sys} is out of the coverage area,
and then its neighbor, F-UE $2$, is selected
to deliver the data traffic.
F-UE $3$ acts as a relay for the data traffic
from traditional UE $1$ to the F-AP, since the maximum transmission power of traditional UE $1$ is limited.
Thus in addition to uploading $R_{th}$ bits at slot $t$ to guarantee its own rate requirement,
F-UE $j$ needs to relay the traffic of other UEs which
are received at the last slot.
The bit rate requirement of F-UE $j$ at slot $t$ is $\sum\limits_{ k=1}^{K_0+K_1} \mathbbm{1}\{\sum\limits_{n=1 }^N s_{k,j,n}^{TX}(t) \geq 1\}R_k(t-1) +R_{th}$, where $\mathbbm{1}\{\sum\limits_{n=1 }^N s_{k,j,n}^{TX}(t) \geq 1\}$ is an indicator function that equals to $1$
when $\sum\limits_{n=1 }^N s_{k,j,n}^{TX}(t) \geq 1$ holds and equals to $0$ otherwise.

\subsection{The computing model}

Computing resource provision in the
BBU pool and F-APs plays a key role in boosting the potential of F-RANs.
As it is shown in the aforementioned communication model,
there are baseband processing and MMSE detector generation.
In this paper, we construct the computing model which follows that in\textcolor[rgb]{1.00,0.00,0.00}{\cite{YLao}}
and corresponding details are as follows.

\begin{itemize}
\item For baseband processing, it consists of inverse fast fourier transform (IFFT), demodulation and decoding.
The IFFT consumes constant computing resource, which is assumed as $C_{cons}$,
while the computing resource required by demodulation and decoding
is approximated as $\mu_1 R_{k}(t)$.

\item For MMSE detector generation, the computational complexity depends on the number of antennas.
Taking the case of $s_{k,0,n}^{TX}(t)=1$ as an example, we assume that the computing resource consumed by the calculation
of $\textbf{v}_{k,0,n}(t)$ is $\mu_0 L_1^3$.
\end{itemize}

Overall, computing resource consumption for UE $k$ are modeled as
\begin{equation}
\label{CPUconsump}
\begin{split}
C_k(t) = &\mu_0 \sum\limits_{n=1 }^N(\sum\limits_{m = 1}^{M_0}s_{k,m,n}^{TX}(t)L_0^3 +s_{k,0,n}^{TX}(t)L_1^3 )+  \\
&\mu_1 R_{k}(t) + C_{cons}, k \in \mathcal{K}_0 \cup \mathcal{K}_1
\end{split}
\end{equation}
where $\mu_0$ and $\mu_1$ are the slopes.
Considering the limited computing resource at F-APs,
the number of UEs accessing F-APs should be under a threshold. Suppose
$D_{m}^{CPU}$ is the computing resource available at F-AP $m$,
we have the following constraint on computing resource consumption.
\begin{equation}
\begin{split}
C1:&D_{m}^{CPU} \geq \sum\limits_{k = 1}^{K_0+K_1} \mathbbm{1}\{\sum\limits_{n=1 }^N s_{k,m,n}^{TX}(t) \geq 1\}C_k(t), \\
&m \in \{0,1,2,...,M_0\}.
\end{split}
\end{equation}

According to the computing model (\ref{CPUconsump}), UEs will consume more computing resource in C-RAN mode than F-AP
mode, since there are more antennas utilized ($L_0 < L_1$). Moreover, there is no computing resource consumption for
the UEs choosing D2D mode.

\section{Problem formulation and Lyapunov Optimization}

In this section,
the concerned optimization problem is
presented at first.
Then with the Lyapunov framework,
the original stochastic problem is reformulated as a deterministic problem at each slot.

\subsection{Problem formulation}

For the concerned
uplink F-RAN, the system power consumption is incurred by
fronthaul transmission and wireless
transmission, which is given by
\begin{equation}
\label{PDef}
\begin{split}
P(t)= &\sum\limits_{i = 1}^{K_0}\sum\limits_{n=1 }^N \frac{1}{\eta_0}P_{i,n}(t)+
\sum\limits_{j = 1}^{K_1}\sum\limits_{n=1 }^N \frac{1}{\eta_1}P_{j,n}(t)\\
&+ \sum\limits_{k = 1}^{K_0+K_1} \sum\limits_{n=1 }^N s_{k,0,n}^{TX}(t) P^{fronthaul},
\end{split}
\end{equation}
where $\eta_0$ and $\eta_1$ are the efficiencies of the power amplifier at each traditional UE and F-UE, respectively,
$P^{fronthaul}$ is the constant power consumption caused by fronthaul transmission.

Despite the mean-rate stable constraint defined in C0
and computing resource constraint defined in C1,
there are also performance constraints to be considered.
As stated in following C2 and C3, the rate of traditional UE $i$ should be larger than its threshold $R_i^{min}$, while
for an F-UE $j$, its rate has to be large enough to upload all the bits in its buffer.
\begin{equation}
\begin{split}
&C2: R_{i}(t) \geqslant R_i^{min}, i\in \mathcal{K}_0, \\
&C3: R_{j}(t) \geqslant \sum\limits_{ k=1}^{K_0+K_1} \mathbbm{1}\{\sum\limits_{n=1 }^N s_{k,j,n}^{TX}(t) \geq 1\}R_k(t-1) \\
&\quad \quad \quad \quad \quad \quad +R_{th}, j\in \mathcal{K}_1.
\end{split}
\end{equation}

To upload traditional UE's bits and maintain the required rate for F-UEs,
a decision on mode selection should be properly made.
Although offloading all the uploaded bits to F-APs
can reduce system power consumption, computing resource at F-APs are limited.
In this paper, our aim is to perform efficient
mode selection and resource allocation, which are described by a tuple $\{s_{k,m,n}^{TX}(t), P_{k,n}(t)\}$.
Combining the constraints
and performance requirements, we formulate the system power optimization
problem as below.
\begin{equation}
\label{Original}
\begin{split}
\min\limits_{\{s_{k,m,n}^{TX}(t), P_{k,n}(t)\}}  \bar{P}= \lim_{T \rightarrow \infty}\frac{1}{T}\sum\limits_{t = 0}^{T-1}  \mathbb{E} \left\{P(t) \right\}
\end{split}
\end{equation}
subjects to
\begin{equation}
\begin{split}
& C0,C1,C2,C3, \\
&C4: P_{k,n}(t) \leq \mathbbm{1}\{\sum\limits_{m = 0}^{M_0+K_1}  s_{k,m,n}^{TX}(t) = 1\} P_{k,n}^{max}, \forall k,n, \\
&C5:  s_{k,m,n}^{TX}(t) \in \{0,1\}, \forall k,m,n,\\
&C6: \sum\limits_{m = 0}^{M_0+K_1}  s_{k,m,n}^{TX}(t)\in \{0,1\}, \forall k,n,\\
&C7: \sum\limits_{m = 0}^{M_0+K_1}\sum\limits_{n=1 }^N  s_{k,m,n}^{TX}(t) \in \{0,1\}, \forall k,\nonumber
\end{split}
\end{equation}
where C0 is to achieve a stable queue backlog for each traditional UE,
C1 is the computing resource constraint,
C2 and C3 are to satisfy the rate requirement for traditional UEs and F-UEs, respectively,
and C4 means if subchannel $n$ is not allocated to UE $k \in \mathcal{K}_0 \cup \mathcal{K}_1$,
the transmission power $P_{k,n}(t)$ has to be 0 and limited by the maximum transmission power $P_{k,n}^{max}$ otherwise.
C5 is the communication mode selection constraint,
C6 implies that at most one mode can be selected by UE $k$ on subchannel $n$, and
C7 means at most 1 subchannel can be allocated to UE $k$.

Solving problem (\ref{Original}) is difficult due to the following
reasons.
First, the problem with aforementioned constraints is a nonlinear optimization problem and falls within the category of
mixed integer programming. Traditional methods like branch-and-bound and genetic algorithms
that can be applied are centralized and will result in high complexity.
Second, the scale of the problem will increase
as the number of traditional UEs/F-UEs grows.
Third, the problem includes future
information like bit rates and queue backlog,
which vary over time and are hard to precisely predict. How to make decisions on $\{s_{k,m,n}^{TX}(t), P_{k,n}(t)\}$ to
adapt to dynamic traffic is of great challenge.

\subsection{General Lyapunov optimization}

Fortunately, with Lyapunov optimization\textcolor[rgb]{1.00,0.00,0.00}{\cite{II:R1}},
the original optimization problem with
the time-averaged constraints C0 can be transformed into a queue mean-rate stable problem,
which can be solved only based on the observed channel state information and queue backlogs at each time slot.
Let ${\bf{Q }}(t) = \{Q_{i}(t)\}$ define queue backlog
set. Taking advantage of Lyapunov optimization, a Lyapunov function is defined
as a scalar metric of queue congestion:
\begin{equation}
L\left( {\bf{Q }}(t) \right)  \triangleq \frac{1}{2} \sum\limits_{i = 1}^{{K_0}}
Q_i^2 (t).
\end{equation}

Then the Lyapunov drift is defined, which pushes the queue backlog to a
lower congestion state and keeps queues stable,
\begin{equation}
\label{eq:II10}
\begin{split}
\Delta \left( {\bf{Q }}(t)\right) \buildrel
\Delta \over =  \mathbb{E} \left\{ L\left( {\bf{Q }}(t+1) \right) -L\left( {\bf{Q }}(t) \right)| {{\bf{Q }}(t)}  \right\}.
\end{split}
\end{equation}

To combine the queue backlog and system power consumption,
the drift-plus-penalty $
\Delta\! \left( {{\bf{Q }}(t)} \! \right)\! +\!
V\!\mathbb{E} \left\{ {P(t)|{\bf{Q }}(t)} \right\}$ is defined,
where $V$ is a non-negative parameter controlling the tradeoff
between the average system power and the average queue delay.
Suppose that the expectation of $P(t)$
is deterministically bounded by finite constants $P_{min}, P_{max}$,
i.e., $P_{min} \leq\mathbb{E} \left\{ {P(t)} \right\}\leq P_{max}$.
Let $P^*$ denote the theoretical optimal value of (\ref{Original}), and then the relationship
between the drift-plus-penalty function and C0 is established in Theorem 1,

\begin{theorem}
\label{LyaOP}
\emph{(Lyapunov optimization)}. Suppose there exist
positive constants $B$, $\epsilon$ and $V$ such that for all slots $t$
and all possible ${\bf{Q }}(t)$, the
drift-plus-penalty function satisfies:
\begin{equation}
\label{eq:II121}
\begin{split}
\Delta \left( {{\bf{Q }}\left( t \right)}
\right) +V\!\mathbb{E} \left\{ {P(t)|{\bf{Q }}(t)} \right\}
\le B +V\!P^* -\! \epsilon\sum\limits_{i = 1}^{{K_0}} {{Q_i}(t)}.
\end{split}
\end{equation}
Then C0 is satisfied and the average system power meets
\begin{equation}
\label{eq:IV25}
\overline{P} = \lim_{T \rightarrow \infty} \frac{1}{T}\sum\limits_{t = 0}^{T-1}  \mathbb{E} \left\{P(t) \right\} \leq {P}^{*} + \frac{B}{V}.
\end{equation}
The average
queue delay is defined as the average length of all queues, which satisfies
\begin{equation}
\label{eq:IV30}
\overline{Q} = \mathop {\lim }\limits_{T \to \infty }  \frac{1}{T}\sum\limits_{t = 0}^{T-1} {\sum\limits_{i = 1}^{{K_0}} {\mathbb{E}\left\{ {{Q_i}(t )} \right\}} } \le \frac{{B \! + \!V \left( { P^{*}-P_{min} }\! \right)}}{\epsilon }.
\end{equation}
\end{theorem}

\begin{IEEEproof}
Since (\ref{eq:II121}) holds for any slot, we can take expectations of both sides
and we have
\begin{equation}
\begin{split}
&\mathbb{E}\left\{ {L({\bf{Q }}(t+1))} \right\} - \mathbb{E}\left\{ {L({\bf{Q }}(t))} \right\}
+ V{\mathbb{E}\left\{ {P(t)} \right\}} \\
 \le & B +VP^*- {\sum\limits_{i = 1}^{{K_0}} {\epsilon {Q_i}(t)} }. \nonumber
\end{split}
\end{equation}

Sum over $t \in \{0,1,2,\cdots,T-1 \}$ and using the law of telescoping sums, it yields
\begin{equation}
\label{eq:APP_B3}
\begin{split}
&\mathbb{E}\left\{ {L({\bf{Q }}(T))} \right\} - \mathbb{E}\left\{ {L({\bf{Q }}(0))} \right\} + V\sum\limits_{t = 0}^{T-1} {\mathbb{E}\left\{ {P(t)} \right\}}\\
 \le & BT +VTP^*- \sum\limits_{t = 0}^{T-1} {\sum\limits_{i = 1}^{{K_0}} {\epsilon {Q_i}(t)} }.
\end{split}
\end{equation}

Based on the fact that $Q_i(t)\geq 0, P_{min} \leq\mathbb{E} \left\{ {P(t)} \right\}$ for all $t$, we rearrange (\ref{eq:APP_B3}) to obtain
yields
\begin{equation}
\begin{split}
\mathbb{E}\left\{ {L({\bf{Q }}(T))} \right\} -\mathbb{E}\left\{ {L({\bf{Q }}(0))} \right\} + VTP_{min}\le BT +VT{P}^{*}, \nonumber
\end{split}
\end{equation}
which could be furthermore rearranged according to definition of Lyapunov function
\begin{equation}
\label{eq:APP_B4}
\begin{split}
\mathbb{E}\left\{ Q_i^2 (T) \right\} \le 2\mathbb{E}\left\{ {L({\bf{Q }}(0))} \right\} + 2BT +2VT{P}^{*}- 2VTP_{min}.
\end{split}
\end{equation}

Note that $\{\mathbb{E}\left\{| Q_i(T) |\right\}\}^2 \le \mathbb{E}\left\{ Q_i^2 (T) \right\}$ holds for any $T$,
we have
\begin{equation}
\mathbb{E}\left\{| Q_i(T) |\right\} \le \sqrt{2\mathbb{E}\left\{ {L({\bf{Q }}(0))} \right\} + 2BT +2VT{P}^{*}- 2VTP_{min}}.
\end{equation}

Dividing both sides by $T$ and taking the limit as $T \to \infty $, we have
\begin{equation}
\lim_{T \rightarrow \infty} \frac{\mathbb{E}\{|Q_{i}(T)|\}}{T} =0.
\end{equation}

According to Definition 1, the queue of traditional UE $i$ is mean-rate stable.
A similar proof can be applied to the queues of other traditional UEs,
which indicates constraint $C0$ is satisfied.

Moreover,
the following inequality is obtained by rearranging the terms in (\ref{eq:APP_B3})
\begin{equation}
\label{eq:APP_C1}
\begin{split}
 V\sum\limits_{t = 0}^{T-1} {\mathbb{E}\left\{ {P(t)} \right\}}
\leq BT + VT{P}^{*} +  \mathbb{E}\left\{ {L({\bf{Q }}(0))} \right\},
\end{split}
\end{equation}
with some non-negative terms neglected when appropriate.
Dividing both sides of (\ref{eq:APP_C1}) by $VT$ and taking the limit as $T \to \infty $,  the inequality (\ref{eq:IV25}) is obtained
based on the fact that $\mathbb{E}\left\{L({\bf{Q }}(0))\right\} < \infty$.

Similarly, inequality (\ref{eq:APP_B3}) can also be re-written as
\begin{equation}
\label{eq:APP_D1}
\begin{split}
\sum\limits_{t = 0}^{T-1} {\sum\limits_{i = 1}^{{K_0}} {\epsilon {Q_i}(t)} } \le& BT +VT{P}^{*} - \mathbb{E}\left\{ {L({\bf{Q }}(T))} \right\} \\
&+ \mathbb{E}\left\{ {L({\bf{Q }}(0))} \right\} - V\sum\limits_{t = 0}^{T-1} {\mathbb{E}\left\{ {P(t)} \right\}}\\
\le& BT +VT{P}^{*}-\mathbb{E}\left\{ {L({\bf{Q }}(T))} \right\} \\
&+ \mathbb{E}\left\{ {L({\bf{Q }}(0))} \right\}- VTP_{min} .
\end{split}
\end{equation}

Dividing (\ref{eq:APP_D1}) by $\epsilon T$ and taking the limit as $T \to \infty$, inequality (\ref{eq:IV30}) is obtained according to the fact that $\mathbb{E}\left\{L({\bf{Q
}}(T))\right\} < \infty$.

\end{IEEEproof}

\emph{Theorem \ref{LyaOP}} suggests that by adjusting the value of parameter $V$,
a near-to-optimal solution can be obtained which provides
an average system power arbitrarily close to the optimum ${P}^{*}$.
Moreover, it is also shown that there exists
an $[\mathcal{O}(1/V),\mathcal{O}(V)]$ tradeoff between the average
system power and the average queue delay. With an increase of parameter
$V$, the achieved system power consumption becomes lower at the cost
of incurring a larger queuing delay. Therefore,
a larger $V$ is suitable for the delay tolerable UEs to obtain the required performance.

Instead of minimizing the drift-plus-penalty directly,
we aim to push the drift-plus-penalty's upper bound to its minimum.
Based on the queue dynamics of ${\bf{Q }}(t)$ and the definition
of Lyapunov drift in (\ref{eq:II10}), the following lemma holds
for the upper bound of drift-plus-penalty.
\begin{lemma}\emph{(Upper bound of Lyapunov drift-plus-penalty)}. At any time slot $t$, with the observed queue state ${\bf{Q }}(t)$
and parameter $V$, there exists an upper bound for
the drift-plus-penalty under any control policy:
\begin{equation}
\label{eq:II12}
\begin{split}
&\Delta \left( {{\bf{Q }}\left( t \right)}
\right) +V\!\mathbb{E} \left\{ {P(t)|{\bf{Q }}(t)} \right\} \\
\le &B +V\!\mathbb{E} \left\{ {P(t)|{\bf{Q }}(t)} \right\} -\! \sum\limits_{i = 1}^{{K_0}} {{Q_i}(t)
\mathbb{E} \left\{ {{R_i}(t) \!-\! {{A}_i}(t)} | {{\bf{Q }}(t)}\right\}}.
\end{split}
\end{equation}
where $B>0$ is a finite constant which is larger than $\frac{1}{2}\sum\limits_{i = 1}^{{K_0}} \mathbb{E}{\left\{ {R_i^2(t) +
{A}_i^2(t)}\!| {{\bf{Q }}(t)} \right\}}$ for any $t$.
\end{lemma}

\begin{IEEEproof}
Squaring both sides of (\ref{eq:I3}) and combining
the inequality $(\max\{Q_{i}(t)-R_{i}(t),0\})^2 \leq (Q_{i}(t)-R_{i}(t))^2$, the following inequality
can be obtained
\begin{equation}
\label{eq:APP_A1}
\begin{split}
Q_{i}^2(t+1)= & \max\{Q_{i}(t)-R_{i}(t),0\}^2+ {A}_i^2(t) \\
&+ 2{A}_i(t)\max\{Q_{i}(t)-R_{i}(t),0\}\\
\leq& (Q_{i}(t)-R_{i}(t))^2+ {A}_i^2(t)\\
&+ 2{A}_i(t)\max\{Q_{i}(t)-R_{i}(t),0\}\\
\leq& Q_{i}^2(t) +R_{i}(t)^2+ {A}_i^2(t)\\
&- 2Q_{i}(t)R_{i}(t)+ 2{A}_i(t)Q_{i}(t)
\end{split}
\end{equation}

Summing (\ref{eq:APP_A1}) over $i \in \{1,2,\cdots,K_0 \}$, we obtain
\begin{equation}
\label{eq:APP_A2}
\begin{split}
 &L\left( {\bf{Q }}(t+1) \right) -L\left( {\bf{Q }}(t) \right)\\
\le&
\frac{1}{2}\sum\limits_{i = 1}^{{K_0}} {\left\{ {R_i^2(t) +
{A}_i^2(t)}\! \right\}}\!\! -\! \sum\limits_{i = 1}^{{K_0}} {{Q_i}(t)
\left\{ {{R_i}(t) \!-\! {{A}_i}(t)} \right\}}. \nonumber
\end{split}
\end{equation}

Taking conditional expectations of both sides, we have
\begin{equation}
\begin{split}
\Delta \left( {{\bf{Q }}\left( t \right)}
\right)\le &
\frac{1}{2}\sum\limits_{i = 1}^{{K_0}} \mathbb{E}{\left\{ {R_i^2(t) +
{A}_i^2(t)}\!| {{\bf{Q }}(t)} \right\}}\\
& -\sum\limits_{i = 1}^{{K_0}} {{Q_i}(t)
\mathbb{E} \left\{ {{R_i}(t) \!-\! {{A}_i}(t)} | {{\bf{Q }}(t)}\right\}} \\
\le &
B -\sum\limits_{i = 1}^{{K_0}} {{Q_i}(t)
\mathbb{E} \left\{ {{R_i}(t) \!-\! {{A}_i}(t)} | {{\bf{Q }}(t)}\right\}}.
\end{split}
\end{equation}
By adding $V\!\mathbb{E} \left\{ {P(t)|{\bf{Q }}(t)} \right\}$, we have (\ref{eq:II12}).

\end{IEEEproof}

Based on the concept of opportunistically
minimizing an expectation,
the policy that minimizes $\mathbb{E} \left\{ {P(t)|{\bf{Q }}(t)} \right\}$ is the one that
minimizes $P(t)$ with the observation of ${\bf{Q }}(t)$ during each slot. Since neither
$Q_i(t)A_i(t)$ nor $B$ in (\ref{eq:II12}) will be affected by the policy at
slot $t$, the upper bound minimization for
the drift-plus-penalty can be accomplished by solving
the following deterministic problem at slot $t$:
\begin{equation}
\label{deter14}
\begin{split}
\mathop {\min }\limits_{\{s_{k,m,n}^{TX}, P_{k,n}\} } &
VP -\sum\limits_{i = 1}^{K_0} {Q_i}R_i \\
s.t. \quad &C1 \sim C7.
\end{split}
\end{equation}
As it is shown in (\ref{deter14}), the power-minus-rate function
as an optimization target is not convex on either
variable $s_{k,m,n}^{TX}$ or variable $P_{k,n}$.


\section{Solution for Orthogonal and Multiplexed Subchannel Strategies}

The non-convex problem (\ref{deter14}),
which includes integer variables $\{s_{k,m,n}^{TX}\}$
and continuous variables $\{ P_{k,n}\}$, is hard to be solved.
Although methods like branch-and-bound and genetic algorithms
can be utilized to solve the integer parts, these existing solutions require a huge complexity
when simultaneously considering all traditional UEs, F-UEs, F-APs and RRHs.
Moreover, the residual part of the
problem (\ref{deter14}) is still non-convex, because
the rate term $R_i$ in the power-minus-rate function depends on the
transmission power $\{P_{k,n}\}$ of traditional UEs and F-UEs using the same subchannel $n$.

In this section, we consider the mode selection and resource allocation under orthogonal and multiplexed subchannel strategies.
To overcome the above challenges,
a centralized approach based on Q-learning and softmax decision-making is proposed for the orthogonal subchannel strategy.
For the multiplexed subchannel strategy, limitations on the subchannel allocation are relaxed. In this case, a distributed approach is developed, where each traditional UE or F-UE needs to consider only its own mode
selection possibilities.

\subsection{Centralized RL-based solution for the orthogonal subchannel strategy}

A centralized approach for mode selection is proposed based on Q-learning.
In particular, the definition of states in Q-learning is related to
current mode selection of UEs.
To decrease the dimensions of the Q table,
the state
is ${\bf{s}}=\{k_0,s_k |k = 1,2,...,K_0+K_1\}$,
in which $k_0$ implies that during the current iteration,
only UE $k_0$ would reselect a mode
according to the action,
and the element $s_k =n+mN$ denotes that subchannel $n$
has been allocated to UE $k$ connecting to F-AP $m$ (namely $ s_{k,m,n}^{TX} =1$).
Considering constraints C5$\sim$C7,
we define the action as $a=n+mN$.
With action $a$ selected, the element $k_0$ and corresponding $s_{k_0}$ in state ${\bf{s}}$
change and the current state transits to the next state.

The Q-value in the Q-learning is
defined as the discounted accumulative reward and starts at a
tuple of a state and an action, which is updated as follows
\begin{equation}
\label{updateQ1}
\begin{split}
Q_{k,m,n} &\leftarrow (1-\alpha)Q_{k,m,n} + \alpha W_{k,m,n} ,
\end{split}
\end{equation}
where $\alpha \in (0, 1)$ is the learning rate, and $W_{k,m,n}$
is the reward resulting from taking action $a$.
Note that in the orthogonal subchannel strategy,
subchannel, for example $n^*$ can not be shared among UEs.
Hence in given state ${\bf{s}}$, there is an element $s_{k'}$
being $n^*+mN(m \in \{0,1,...,M_0+K_1\})$.
If the action is chosen and $a = n^*+m'N(m' \in \{0,1,...,M_0+K_1\})$,
the reward has to be 0 ($W_{k,m,n^*}=0$). Otherwise,
the value of reward $W_{k,m,n}$ is defined as a value between $0$ and $1$ that decreases when the power-minus-rate increases:

\begin{equation}\label{rewardDef}
W_{k,m,n} \!= \!\left\{\! {\begin{array}{*{20}{c}}
\!1-\! \frac{V_0P_{k,n} +{s}_{k,0,n}^{TX}VP^{fronthaul}- {Q_k}R_{k,m,n}}{V_0P_{k,n}^{max}+{s}_{k,0,n}^{TX}VP^{fronthaul}- {Q_k}R_k^{min}},\! & k \!\in \!\mathcal{K}_0,  \\
\!1 -\! \frac{V_1P_{k,n} +{s}_{k,0,n}^{TX}VP^{fronthaul}}{V_1P_{k,n}^{max} +{s}_{k,0,n}^{TX}VP^{fronthaul}}, \!&  k \!\in \!\mathcal{K}_1,
\end{array}} \right.
\end{equation}
where $V_0 = \frac{V}{\eta_0}$ and $V_1 = \frac{V}{\eta_1}$. Note that the reward function is defined according to the UE's performance requirement. Since the mean-rate stable is considered only for each traditional UE, the reward function of UE is different from F-UE's.

Here, the softmax selection policy\textcolor[rgb]{1.00,0.00,0.00}{\cite{RRBook}} is used
to determine the communication mode.
The probability $Pr_{k,m,n}$ of UE $k$ selecting F-AP $m$ on subchannel $n$ is calculated as
\begin{equation}
\label{soft}
\begin{split}
Pr_{k,m,n} = \frac{ e^{\frac{Q_{k,m,n}}{\tau}} } { \sum\limits_{m'=0}^{M_0+K_1}\sum\limits_{n'=1}^{N} e^{\frac{Q_{k,m',n'}}{\tau}} },  k \in \mathcal{K}_0 \cup \mathcal{K}_1,
\end{split}
\end{equation}
where $\tau = \tau_0 / \log(1+ t_{epi})$ is the temperature parameter.
At the beginning, the temperature parameter is high, which
leads to a nearly equiprobable selection among the different modes.
As the episode $t_{epi}$ increases,
the value of the temperature parameter decreases and greater difference in
selection probabilities $\{Pr_{k,m,n}\}$ occurs.
The larger the estimated value of $Q_{k,m,n}$ is, the higher the probability
$Pr_{k,m,n}$ is.

After $\{s_{k,m,n}^{TX}\}$ are identified via Q-learning, problem (\ref{deter14}) is simplified into the following problem.
\begin{equation}
\label{Ortho}
\begin{split}
\mathop {\min } \limits_{\{P_{k,n}\} } \quad &
\sum\limits_{i = 1}^{K_0}\sum\limits_{n=1 }^N {V}_{0}P_{i,n}+
\sum\limits_{j = 1}^{K_1}\sum\limits_{n=1 }^N {V}_{1}P_{j,n} -\sum\limits_{i = 1}^{K_0} {Q_i}R_i \\
s.t.\quad & C1\sim C4.
\end{split}
\end{equation}
Since subchannel $n$ is allocated to at most one UE in the orthogonal subchannel allocation strategy,
the interference part $Int_{k,m,n}$ in (\ref{RateDef}) equals to $0$ and the rate $R_k$ in ($\ref{Ortho}$) is convex and monotonically increases with the power $P_{k,n}$.
Suppose $\{P_{k,n}^*\}$ is the extreme point of the targeted convex function.
When $\{P_{k,n}^*\}$ is in the feasible region defined by C1 $\sim$ C4,
$\{P_{k,n}^*\}$ is the optimal solution of problem (\ref{Ortho}).
When $\{P_{k,n}^*\}$ is not in the feasible region, we can find the optimal
solution by the following iterative methods:

\begin{algorithm}[htb]
\caption{An iterative method to find the optimal solution of problem (\ref{Ortho}).}\label{alg:itt}
\begin{algorithmic}[1]
\STATE Derive the partial derivative of the targeted optimization function in (\ref{Ortho});
\STATE Find the extreme point $\{P_{k,n}^*\}$ of the targeted convex function.
\STATE Initialize $\{P_{k,n}\} =\{P_{k,n}^*\}$ and define a fixed step $\triangle P$;
\REPEAT
\STATE With $P_{k,n}$ fixed, calculate the partial derivative $f'(P_{k,n})$;
\STATE Find the minimal one $k^* = \arg\min_k f'(P_{k,n})$;
\STATE Update $P_{k^*,n} = P_{k^*,n} - \triangle P$;
\UNTIL $\{P_{k,n}\}$ is in the feasible region.
\end{algorithmic}
\end{algorithm}

\subsection{Distributed RL-based solution for multiplexed subchannel allocation strategy}

In the multiplexed subchannel allocation strategy, a distributed RL-based approach is proposed,
in which UEs autonomously select their communication modes. The main advantage of using distributed approaches
is that they allow for a reduction in complexity since each UE needs to consider only its own
selection possibilities.
Note that the size of Q-table can be decreased
by only considering the neighbor nodes of UE $k$,
which makes the storage of Q-table affordable for each UE.

Whenever RRHs($m=0$), an F-AP($m=\{1,2,...,M_0\}$) or an F-UE($m=\{M_0+1,M_0+2,...,M_0+K_1\}$) and subchannel $n$
has been selected by UE $k$, the
value of $Q_{k,m,n}$ is updated as (\ref{updateQ1}).
Unlike the special case $W_{k,m,n}= 0$ in the orthogonal subchannel allocation strategy, a subchannel can be shared among multiple UEs in the multiplexed subchannel allocation strategy.
We have to consider the following cases in which $W_{k,m,n}$ are supposed to be 0:
1) An excessive load occurs in F-AP $m$ and there is no enough computing resource for the connected UEs, meaning that
constraint C1 is not fulfilled;
2) The propagation conditions in the selected mode do not
allow guaranteeing the traditional UE's rate requirement, meaning that constraint C2 is not satisfied;
3) The propagation conditions in the selected mode do not
allow achieving the desired rate of F-UE, meaning that constraints C3 is not satisfied.
If constraints C1, C2 and C3 are satisfied,
we have the same definition on the reward $W_{k,m,n} $ as in (\ref{rewardDef}).
By defining a reward with C1 $\sim$ C3 and the power-minus-rate function considered,
the reward $W_{k,m,n}$ reflects the degree
of fulfillment of the optimization target and the constraints.

Based on communication modes $\{s_{k,m,n}^{TX}\}$ output by distributed Q-learning,
there is a fixed one-to-one mapping between $k$ and $\{m,n\}$ due to constraints C5$\sim$C7.
Define the corresponding
mode selection and subchannel allocation for UE $k$
as $m(k)$ and $n(k)$, respectively.
Note that when subchannel $n(k)$ is used by a single UE,
the interference part is omitted, which makes the problem convex.
When subchannel $n(k)$ is reused, for example by UE $k'$ and $k$, we have
$n(k) = n(k')$.
Problem (\ref{deter14}) can now be simplified into the following problem at subchannel $n(k)$.
\setcounter{equation}{27}
\begin{figure*}[hb]
\hrulefill
\begin{equation}\label{eq:II16}
\begin{aligned}
\mathop {\min } \limits_{\{P_{k,n(k)}\} } \quad &
\sum\limits_{i = 1}^{K_0} {V}_{0}P_{i,n(i)}+
\sum\limits_{j = 1}^{K_1} {V}_{1}P_{j,n(i)} -\sum\limits_{i = 1}^{K_0} {Q_i}R_i \\
s.t.\quad & C1,C4\\
&D2: \sqrt{\sum\limits_{k'=1 }^{K_0+K_1} P_{k',n(k')} \|\textbf{v}_{k,m(k),n(k)}^H\textbf{h}_{k',m(k),n(k)}\|^2+\sigma ^2\|\textbf{v}_{k,m(k),n(k)}\|^2} \\
&\quad \leq  \sqrt{1+\frac{1}{\gamma_k^{QoS}}} {\bf{Re}}\{ \textbf{v}_{k,m(k),n(k)}^H\textbf{h}_{k,m(k),n(k)} \}P_{k,n(k)}^{\frac{1}{2}}, k\in \mathcal{K}_0 \cup \mathcal{K}_1,
\end{aligned}
\end{equation}
\end{figure*}
where $\gamma_k^{QoS}$ is the SINR corresponding
to the desired rate $R_i^{min}$ in C2 and sum rate threshold in the right side of C3.
The second order cone constraint D2 is transformed from C2 and C3 equivalently.

The target function in (\ref{eq:II16}) is non-convex when subchannel $n(k)$ is reused.
Hence, a \emph{C-additive approximation} of the drift-plus-penalty algorithm is presented, the performance of which is
within an additive constant of the infimum.
The definition of \emph{C-additive approximation}\textcolor[rgb]{1.00,0.00,0.00}{\cite{II:R1}} is defined as follows.
\begin{Definition}
\emph{(C-additive approximation)}.
For a given constant $C \geq 0$, a \emph{C-additive approximation}
of the drift-plus-penalty algorithm is to choose an action that
yields a conditional expected value on the right-hand-side of the
drift-plus-penalty under given ${\bf{Q }}(t)$ at time slot
$t$, which is within a constant $C$ from the infimum over all
possible control actions.
\end{Definition}

The \emph{C-additive approximation} of the drift-plus-penalty algorithm
is inspired by the equivalence between
the weighted sum rate maximization and WMMSE\textcolor[rgb]{1.00,0.00,0.00}{\cite{III:R11}} for the MIMO channel, which is
extended to solve problem (\ref{eq:II16}). We state this equivalence as follows.


\begin{proposition}
\emph{(Equivalent WMMSE problem)}.
Problem (\ref{eq:II16}) has the same optimal solution as the following WMMSE problem:
\begin{equation}
\label{eq:II18}
\begin{split}
\mathop {\min } \limits_{\{w_k,u_k,P_{k,n(k)}^{\frac{1}{2}}\}} \quad & \sum\limits_{i = 1}^{{K_0}} {Q_i \left\{ {w_i e_i - \log w_i} \right\}}\\
&   + \sum\limits_{i = 1}^{K_0} {V}_0P_{i,n(i)}+\sum\limits_{j = 1}^{K_1} {V}_1P_{j,n(j)} , \\
s.t.\quad&C1,C4,D2,
\end{split}
\end{equation}
where $w_k$ denotes the mean-square error (MSE) weight for UE $k$, $u_k \in \mathbb{C}$ is a receiver variable, and $e_k$ is the corresponding MSE defined as
\begin{equation}
\label{eq:II19}
\begin{split}
{e_k} \buildrel \Delta \over = &\|u_k \sum\limits_{k'}\textbf{v}_{k,m(k),n(k)}^H\textbf{h}_{k',m(k),n(k)}P_{k',n(k')}^{\frac{1}{2}}\|^2    \\
&- 2{\bf{Re}}\{u_k \textbf{v}_{k,m(k),n(k)}^H\textbf{h}_{k,m(k),n(k)} \}P_{k,n(k)}^{\frac{1}{2}} \\
&  +\sigma ^2 \|u_k \textbf{v}_{k,m(k),n(k)}\|^2 + 1.
\end{split}
\end{equation}
\end{proposition}

Note that WMMSE problem (\ref{eq:II18}) is not jointly convex in
$w_k,u_k$, and $P_{k,n(k)}^{\frac{1}{2}}$ but convex with respect to each of the individual
optimization variables when other individuals are fixed. Hence, the
block coordinate descent (BCD) method is utilized to obtain
a stationary point of problem (\ref{eq:II18}). The BCD method is summarized as follows
and described in \textbf{Algorithm \ref{alg:1}}.

\begin{itemize}
\item The optimal receiver $u_k$ under the fixed $P_{k,n(k)}^{\frac{1}{2}}$ and $w_k$ is
given by
\begin{equation}
\label{eq:II22}
\begin{split}
u_k^{opt} =  &\textbf{v}_{k,m(k),n(k)}^H\textbf{h}_{k,m(k),n(k)} P_{k,n(k)}^{\frac{1}{2}} \\
&  \Big\{
\sum\limits_{k' }\| \textbf{v}_{k,m(k),n(k)}^H\textbf{h}_{k',m(k),n(k)}\|^2   P_{k',n(k')}\\
&  +\sigma ^2 \| \textbf{v}_{k,m(k),n(k)}\|^2 \Big\}^{-1}.
\end{split}
\end{equation}

\item The optimal MSE weight $w_k$ under the fixed $P_{k,n(k)}^{\frac{1}{2}}$ and $u_k$ is
given by
\begin{equation}
\label{eq:II20} w_k^{opt}= \ e_k^{ - 1}.
\end{equation}

\item Note that the optimization problem for finding the optimal transmit power $P_{k,n(k)}$ under the fixed $u_k$ and $w_k$ is
\begin{equation}
\label{eq:II21}
\begin{split}
\mathop {\min } \limits_{\{P_{k,n(k)}^{\frac{1}{2}}\}} \quad & \sum\limits_{i = 1}^{{K_0}} {Q_i {w_i e_i } } + \sum\limits_{i = 1}^{K_0} {V}_0P_{i,n(i)}+
\sum\limits_{j= 1}^{K_1} {V}_1P_{j,n(j)} , \\
s.t. \quad& C4,D2,\\
&D3: s_{m}^{CPU}D \geq \sum\limits_{k = 1}^{K_0+K_1} \mathbbm{1}\{\sum\limits_{n=1 }^N s_{k,m,n}^{TX} \geq 1\}\tilde{C}_k,
\end{split}
\end{equation}
%
which is a second order cone problem and
can be solved efficiently when there is convex region.
Note that the convex region is defined by the constraint
C4, D2 and D3 jointly. In particular, the new constraint D3 is derived from C1.
In constraint C1, the computing resource consumption $C_k(t)$ of UE $k$
is calculated according to the
resource allocation under determined mode selection.
While in the presented BCD method, the resource allocation
is determined in an iterative way. Hence
$\tilde{C}_k$ in D3 is calculated based on the power output by the last iteration.
\end{itemize}

\begin{algorithm}[htb]
\caption{WMMSE algorithm for solving (\ref{eq:II18}).}\label{alg:1}
\begin{algorithmic}[1]
\STATE For each slot $t$, observe the current ${\bf{Q }}(t)$ and $\textbf{h}_{k,m(k),n(k)}$, and then make the queue-aware power allocation according to the following steps:
\STATE Initialize the precision $\kappa$, power $P_{k,n(k)}$ and corresponding power-minus-rate function $PMR$;
\REPEAT
\STATE Update $P_k = P_{k,n(k)}$ and $PMR^* = PMR$;
\STATE With $P_k$ fixed, compute $u_k$ according to (\ref{eq:II22});
\STATE Compute the corresponding MSE $e_k$
according to (\ref{eq:II19}) and set $w_k = e_k^{-1}$;
\STATE Find the optimal value of power $P_{k,n(k)}$ by solving problem (\ref{eq:II21});
\STATE Calculate the corresponding power-minus-rate function $PMR$ in (\ref{eq:II16});
\UNTIL Constraint D3 is not satisfied or $|PMR - PMR^*|\leq \kappa |PMR^*|$;
\STATE Update ${\bf{Q }}(t)$.
\end{algorithmic}
\end{algorithm}

As proven in\textcolor[rgb]{1.00,0.00,0.00}{\cite{III:R11}},
a fixed point of problem (\ref{eq:II18}) will be reached when \textbf{Algorithm 2}
converges, which might not be globally optimal for problem (\ref{eq:II16}) or (\ref{eq:II18}).
To enable a quick convergence, it is critical to choose proper initialization points
with reasonable approaches like the interference alignment initialization.

\textbf{Algorithm 2} is based on the BCD method. In this case, the computational complexity
of Step 4 is $O(K_0+K_1)$.
For Step 5 and 6, the computational complexity is $O(K_0L_0(K_0+K_1))$.
In Step 6, the additional computational complexity to update
all MSE weights $w_k$ is only $O(K_0)$
Step 7
is the largest part of the computational complexity in \textbf{Algorithm 2}.
The total number of variables in the problem is
$(K_0+K_1)L_0$ and the computation complexity of using the CVX
method to solve such an problem is approximately
$O(((K_0+K_1)L_0)^{3.5})$.

\section{Simulation results}
To demonstrate the performance
of the proposed RL-based solutions,
extensive simulation has been conducted.
Assume $L_1 = 10$ RRHs and $M_0 = 3$ F-APs that are deployed in a square
region $1000 m \times 1000 m$.
Each F-AP is equipped with $L_0 = 6$ antennas.
We also assume that the mean arrival rate $\lambda_i$ of each traditional UE is the same.
For each F-UE, the bit rate requirement is $R_{th} = 0.6$ Mbits/slot,
and the bit rate requirement of the traditional UE is set to $R_i^{min} = 0.06$ Mbits/slot.
The pathloss is modeled as $127 + 25 \log_{10}(d)$ with $d$ (km) being
the propagation distance.
The subchannel bandwidth is
180 kHz and the noise power spectral density is $-164$ dBm/Hz.
Each simulation experiment is run for 10000
time slots.
A summarization on the parameters in the simulation are shown in Table I.

\begin{table}[h]
\centering \caption{Simulation Parameters}
\small
\begin{tabular}{|m{4cm}<{\centering}|m{4cm}<{\centering}|}
\hline
 Fronthaul power $P^{fronthaul}$ & $0.35$ W \\
\hline
 Rate threshold $R_{th}, R_i^{min}$ & $0.6, 0.06$ Mbits/slot \\
\hline
 Noise power spectral density & $-164$ dBm/Hz  \\
\hline
 Subchannel bandwidth $W_0$ & $180 $ kHz  \\
\hline
 Power amplifier efficiencies $\eta_0, \eta_1$ & $0.05,0.05$  \\
\hline
 Pathloss model & $127 + 25 \log_{10}(d)(km)$  \\
\hline
\end{tabular}
\label{TableChannel1}
\end{table}

\subsection{The impacts of different parameters}

\begin{figure}[!htb]
\centering
\includegraphics[width=0.45\textwidth]{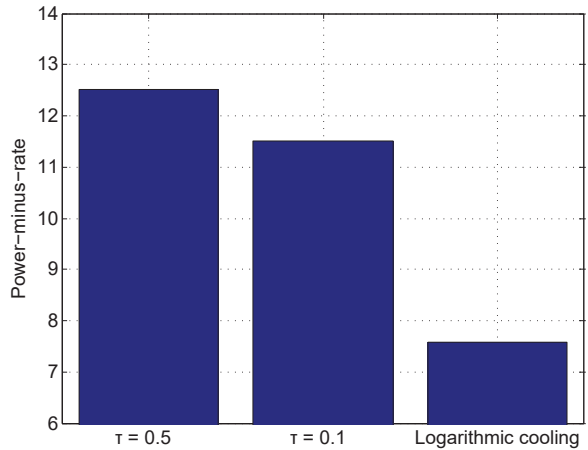}
\caption{The impacts of the temperature parameter $\tau$ in the orthogonal subchannel strategy.}
\label{comp2}
\end{figure}

The impact of temperature parameter $\tau$
on system performance is illustrated in Fig. \ref{comp2}.
It can be observed that a smaller value of
$\tau$ achieves a better performance.
This is because a bigger $\tau$ will lead to a near equal
selection probabilities for different actions, even if
the gap between their Q values becomes large after a period of learning.
It is also shown that the performance
is benefited from a logarithmic decreasing $\tau= \tau_0 / \log(1+ t_{epi})$,
compared with $\tau = 0.1$ and $\tau = 0.5$.
This is because logarithmic decreasing $\tau$ tends to reduce
its value as the episode $t_{epi}$ increases, and therefore,
the best solutions are progressively
selected with higher probability.

\begin{figure}[!htb]
\centering
\includegraphics[width=0.45\textwidth]{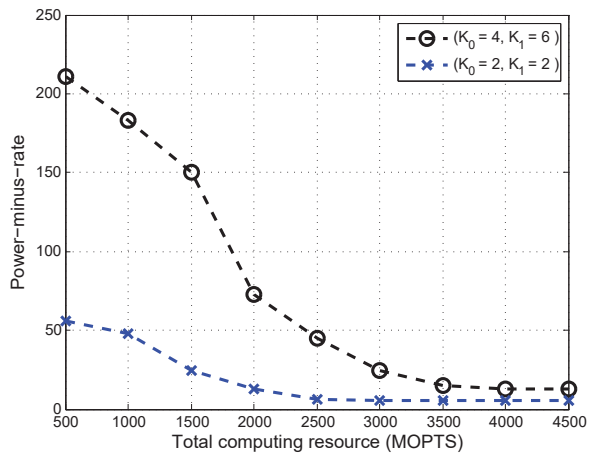}
\caption{Power-minus-rate v.s. total available computing resource
$\sum_m D_{m}^{CPU}$ at slot $t$.}
\label{PowerVSMOPT}
\end{figure}

Fig. \ref{PowerVSMOPT} shows the relationship between the total available computing resource and the power-minus-rate. The total computing resource
is determined by the number of active
processors and computing capability of each processor, the unit of which is
million operations per time slot (MOPTS).
From Fig. \ref{PowerVSMOPT},
it can be seen that when total available computing resource is scarce, the value of target power-minus-rate function will be significantly limited.
With the computing resource increasing,
the power-minus-rate decreases significantly, which may be because of the following reasons.
First, as
the computing resource available increases, more traditional
UEs/F-UEs can be served locally. With more flexible mode selection,
the power-minus-rate can be decreased;
Second, with more computing resource,
UEs/F-UEs used to select D2D mode may choose F-AP mode.
Since the F-UE is free from relaying data,
the power consumption of the F-UE decreased,
which further decreases the power-minus-rate.

\begin{figure}[!htb]
\centering
\subfigure[Average system power $\overline{P}$ v.s. parameter $V$]{
\label{Tradeoff1} 
\includegraphics[width=0.45\textwidth]{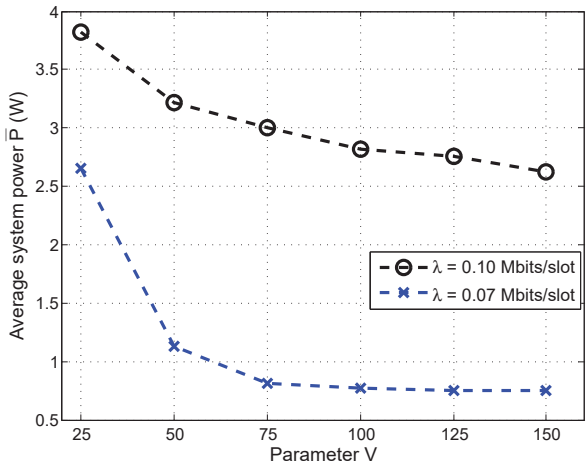}}
\hspace{1in}
\subfigure[Average queue delay $\overline{Q}$ v.s. parameter $V$]{
\label{Tradeoff2} 
\includegraphics[width=0.45\textwidth]{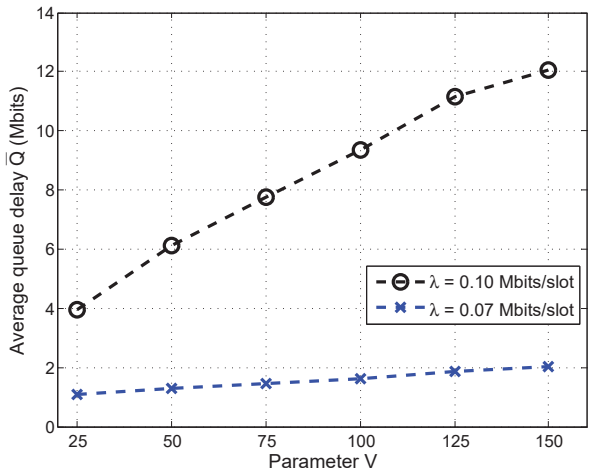}}
\caption{A tradeoff between the average system power $\overline{P}$ and average queue delay $\overline{Q}$, which
can be controlled via the parameter $V$.}
\label{Tradeoff12} 
\end{figure}

Despite the performance at deterministic slot,
we also evaluate the average
queue delay and the average system power.
It is observed in Fig. \ref{Tradeoff12} that when the mean arrival rate $\lambda$
is larger, longer average delay and higher system power will occur.
This can be explained by the fact that more power is needed
to timely transmit larger amount of traffic arrivals. Under
a given mean arrival rate, the average system power
is a monotonically decreasing function on parameter $V$,
which is consistent with \emph{Theorem \ref{LyaOP}}.
As illustrated in Fig. \ref{Tradeoff1}, the decreasing rate of average system
power starts to diminish with excessive increase
of $V$. On the other hand,
a larger $V$ can adversely affect the
delay performance, which leads to higher average queue delay shown in Fig. \ref{Tradeoff2}. This is because that the algorithm with a
larger $V$ will emphasize less on delay performance but more
on the system power performance. Therefore the
parameter $V$ features the tradeoff between power consumption
and delay performance.

\subsection{Performance comparison with benchmarks}

\begin{figure}[!htb]
\centering
\includegraphics[width=0.45\textwidth]{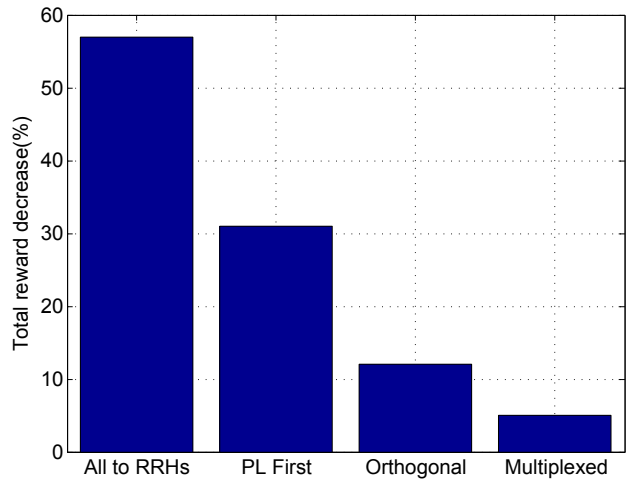} 
\caption{Performance evaluation with respect to the optimum solution,
including the All to RRHs, the PLFirst, and the proposed algorithm under orthogonal and multiplexed subchannel strategy.}
\label{comp1}
\end{figure}

Fig. \ref{comp1} presents an evaluation of the proposed RL-based algorithms.
Two mode selection approaches are included for comparison:
The first one is the approach that all traditional UEs and F-UEs are connected
to the RRHs (denoted as \textquotedblleft All to RRHs\textquotedblright),
where F-AP mode and D2D mode
are not provided and the subchannel is selected randomly; The second one is the
\textquotedblleft PL First\textquotedblright approach in which the traditional UE and F-UE selects an F-AP/RRHs
with the lowest propagation loss.
The performance are compared with respect to the optimum solution.
To behave an exhaustive search to obtain the optimal mode selection,
we consider only $K_0=2$ traditional UEs and $K_1=2$ F-UEs to be served
with a total of $N = 4$ subchannels in this simulation.
It is demonstrated in Fig. \ref{comp1} that
compared with the All to RRHs approach,
the total reward of the proposed RL-based algorithms can be decreased significantly.
This is because that the proposed algorithms take advantages of computing resources
at F-APs, which leads a save on the fronthaul power consumption.
It is also shown that the proposed algorithms outperform than the
PL First approach. Since data relay of F-UEs enables more
UEs and F-UEs served locally and a more efficient mode
selection is achieved via RL.
Note that there are less constraints on the subchannel selections
in the multiplexed subchannel allocation strategy, the performance of the proposed algorithm under multiplexed subchannel allocation strategy is better than the orthogonal strategy.

When the number of traditional UEs, F-UEs, F-APs
and subchannels increases, the number of combinations $\{s_{k,m,n}^{TX}(t)\}$
becomes large dramatically, which makes it unfeasible to obtain the optimum solution
via exhaustive search.
Hence we apply the particle swarm optimization (PSO) approach
as a benchmark. Specially,
$E$ particles
are defined, each of which is with
a corresponding position $\mathbf{x}^{e,0}$
and velocity $\mathbf{v}^{e,0}$.
The position $\mathbf{x}^{e,0}$ of particle $e$ is used to generate
a mode selection, while the velocity $\mathbf{v}^{e,0}$
is used to update the position.
The operation of the PSO
approach at slot $t$ is summarized as follows.

\begin{enumerate}
\item At initialization, a position set $\{\mathbf{x}^{e,0}\}$ and corresponding velocity set
$\{\mathbf{v}^{e,0}\}$ of $E$ particles are randomly generated;

\item At each iteration $u$, the following operators are applied to the particle position updates
to obtain the new position set of particles:

\begin{enumerate}[a)]
    \item For each particle $e$, there is a mapping from $\mathbf{x}^{e,u}=\{x^{e,u}_k | k= 1,2,...,{K_0+K_1}\}$ to $\{s_{k,m,n}^{TX}(t)\}$,
        \begin{equation}
        s_{k,m,n}^{TX}(t)=
        \begin{cases}
        1& \text{if $n+mN=\left \lfloor {x^{e,u}_k} \right \rfloor $,}\\
        0& \text{others,}
        \end{cases}
        \end{equation}
        where $\left \lfloor {x^{e,u}_k} \right \rfloor $ is a floor function that outputs the greatest integer less than or equal to $x^{e,u}_k$.

    \item Given $\{s_{k,m,n}^{TX}(t)\}$, evaluate the fitness value which is defined as follows;
        \begin{equation}
        \begin{split}
        \mathop {\min }\limits_{\{P_{k,n}\} } &
        VP -\sum\limits_{i = 1}^{K_0} {Q_i}R_i  \\
        s.t.\quad & C1\sim C4
        \end{split}
        \end{equation}

    \item According to the fitness value, update the personal best position of each particle $\mathbf{p}^{e,u}$
    and the global best position $\mathbf{g}^u$ during past $u$ iterations;

    \item Update the velocity $\mathbf{v}^{e,u+1}$ and position $\mathbf{x}^{e,u+1}$ of each particle
        \begin{equation}
        \begin{split}
        \mathbf{v}^{e,u+1}&= \!w\mathbf{v}^{e,u} +\!r_1c_1(\mathbf{p}^{e,u}-\!\mathbf{x}^{e,u})+\!r_2c_2(\mathbf{g}^{u}-\!\mathbf{x}^{e,u}),\\
        \mathbf{x}^{e,u+1}&= \mathbf{v}^{e,u+1} +\mathbf{x}^{e,u},
        \end{split}
        \end{equation}
        where $w$ is a weight factor, $r_1,r_2$ are random constants
        to increase search randomness, and $c_1,c_2$ are used to adjust learning maximum step size.
\end{enumerate}

\item Until $U$ iterations, the global best position $\mathbf{g}^{U-1}$ corresponds to the final selection result.
\end{enumerate}

\begin{figure}[!htb]
\centering
\includegraphics[width=0.45\textwidth]{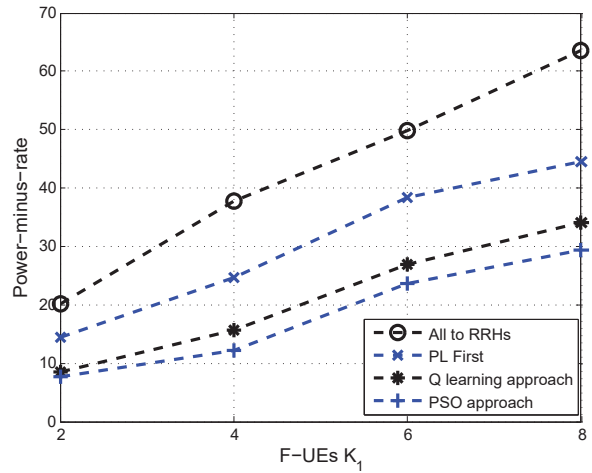}
\caption{The performance of proposed approach at slot $t$ when
increasing the number of F-UEs with multiplexed subchannel allocation strategy.}
\label{compPSO}
\end{figure}

The corresponding evaluation is performed in the scenario with
$M_0 = 6$ F-APs and $N = 6$ subchannels. There are
$K_0=6$ traditional UEs to be served and the multiplexed subchannel strategy is considered.
Fig. \ref{compPSO} shows a comparison between the proposed Q-learning approach
and the PSO approach. It is shown that the proposed Q-learning approach always outperforms the All to RRHs approach
and PL First approach under different number of F-UEs $K_1$. Moreover, the proposed Q-learning approach
achieves similar performance to the PSO approach. Despite close performance,
the Q-learning approach performs much better than the PSO approach in terms of
computational complexity. Specially, to obtain the presented performance result in Fig. \ref{compPSO},
the execution of the PSO approach costs approximately
35 minutes, while the Q-learning approach costs only 2 minutes.

\section{Conclusion and future work}
In this paper, we have investigated the mode selection and resource allocation problem
in a sliced F-RAN.
In particular, network slice instances are constructed
to satisfy specific performance requirements of traditional UEs and F-UEs.
To guarantee the slice isolation,
both orthogonal and multiplexed subchannel allocation strategies are presented.
With performance requirements and limited resources considered,
a system power minimization problem is formulated
and two RL-based approaches are developed.
In the RL-based approaches, an opportunistic mode selection is performed based on the learned policy,
with transmission power of traditional UEs and F-UEs optimized subsequently.
By simulation, benefits of the proposed approaches are
validated and a delay-power tradeoff has been achieved.

There are still some other topics to be researched in the future work.
For example, the extend of our work to other network slicing scenarios like Internet of vehicles.
Key challenges for machine learning and artificial intelligence techniques are
also interesting to be investigated, such as the robustness improvement to model drift and
generalization of algorithms. Besides,
consider the emerging applications and uses cases,
network slicing method in F-RANs should be furthermore explored.
The novel approaches may have to consider signaling overhead, joint allocation
of computing, caching and radio resource. To guarantee differentiated demands,
the system design should address issues like the quality of service, scalability.

%
%
%

%

\begin{IEEEbiography}
[{\includegraphics[width=1in,height=1.25in,clip,keepaspectratio]{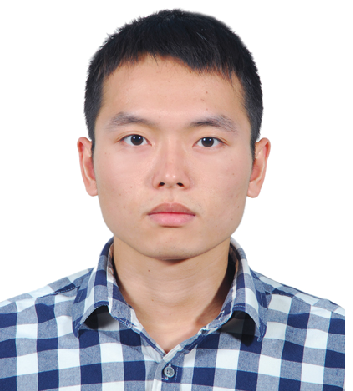}}]{Hongyu Xiang}
is currently pursuing the Ph.D. degree at the Beijing
University of Posts \text{\&} Telecommunications (BUPT). He received
the B.S. degree in communication engineering from the Fudan
University, China, in 2013. His research interests are network slicing, cooperative
radio resource management and collaboration radio signal processing
in large-scale networks like the fog radio access networks
(F-RANs).
\end{IEEEbiography}

\begin{IEEEbiography}
 [{\includegraphics[width=1in,height=1.25in,clip,keepaspectratio]{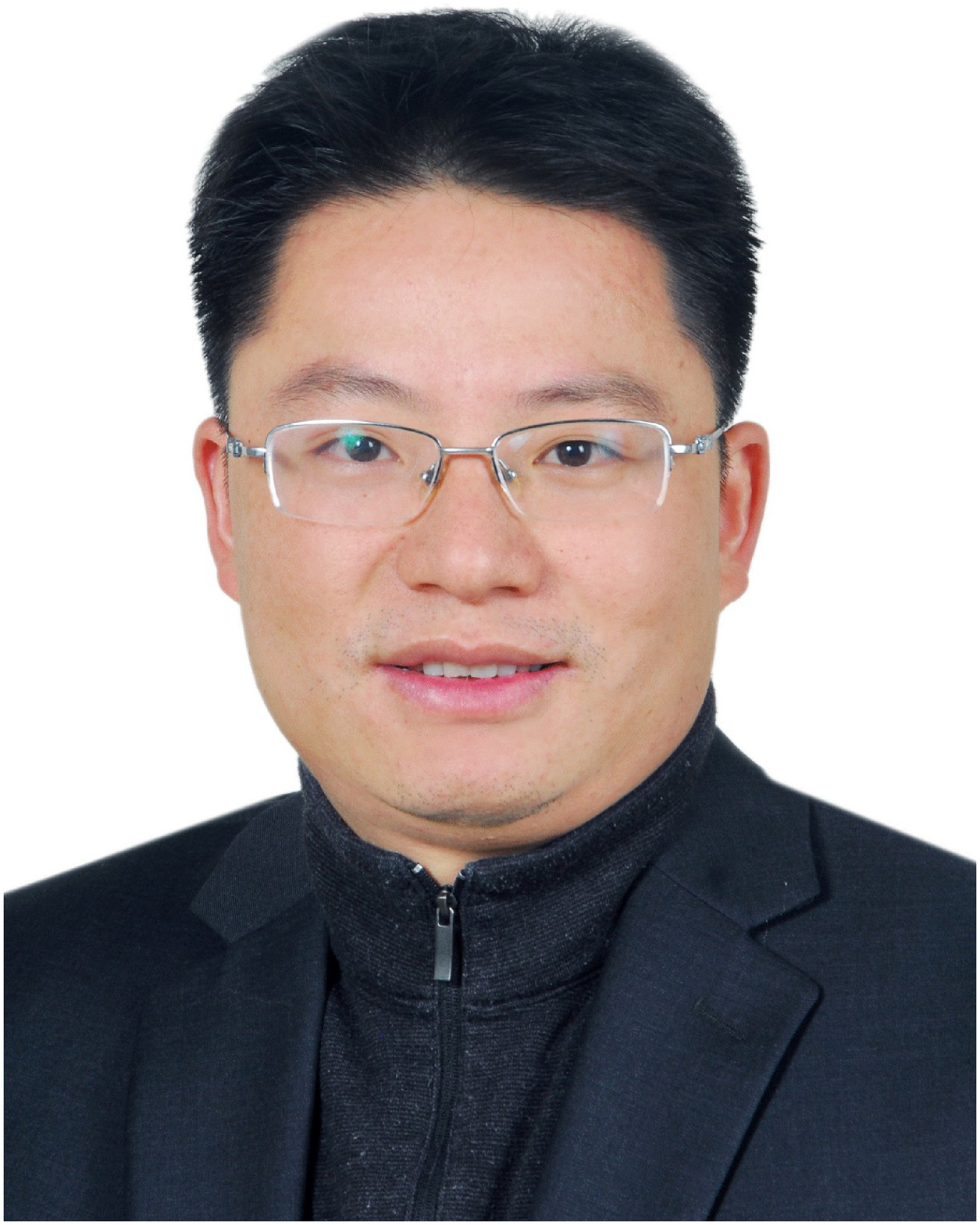}}]{Mugen Peng}
 (M'05-SM'11-F'20) received the
Ph.D. degree in communication and information
systems from the Beijing University of Posts and
Telecommunications (BUPT), Beijing, China, in
2005. Afterward, he joined BUPT, where he has
been a Full Professor with the School of Information
and Communication Engineering since 2012.
In 2014, he was an Academic Visiting Fellow with
Princeton University, Princeton, NJ, USA. He leads
a Research Group focusing on wireless transmission
and networking technologies with the State
Key Laboratory of Networking and Switching Technology, BUPT. He has
authored/coauthored over 90 refereed IEEE journal papers and over 300
conference proceeding papers. Dr. Peng was a recipient of the 2018 Heinrich
Hertz Prize Paper Award, the 2014 IEEE ComSoc AP Outstanding Young
Researcher Award, and the Best Paper Award in the JCN 2016, IEEE WCNC
2015, IEEE GameNets 2014, IEEE CIT 2014, ICCTA 2011, IC-BNMT 2010,
and IET CCWMC 2009. He is on the Editorial/Associate Editorial Board of
the IEEE Communications Magazine, IEEE Access, IET Communications,
IEEE Internet of Things Journal, and China Communications.
\end{IEEEbiography}

\begin{IEEEbiography}
 [{\includegraphics[width=1in,height=1.25in,clip,keepaspectratio]{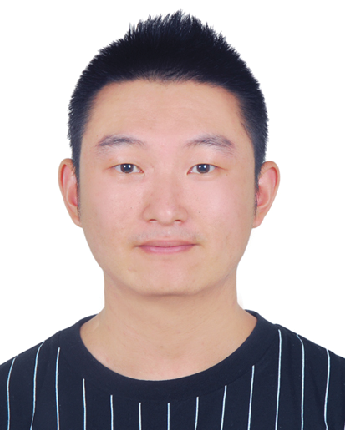}}]{Yaohua Sun}
 received the bachelor's degree (with first class Hons.) in telecommunications engineering (with management) and Phd degree in communication engineering both from Beijing University of Posts and Telecommunications (BUPT), Beijing, China, in 2014 and 2019, respectively. He is currently an assistant professor at the State Key Laboratory of Networking and Switching Technology (SKL-NST), BUPT. His research interests include IoT, edge computing, resource management, (deep) reinforcement learning, network slicing, and fog radio access networks. He was the recipient of the National Scholarship in 2011 and 2017, and he has been reviewers for \emph{IEEE Transactions on Communications},  \emph{IEEE Transactions on Mobile Computing}, \emph{IEEE Systems Journal}, \emph{Journal on Selected Areas in Communications}, \emph{IEEE Communications Magazine},
\emph{IEEE Wireless Communications Magazine}, \emph{IEEE Wireless Communications Letters}, \emph{IEEE Communications Letters}, and \emph{IEEE Internet of Things Journal}.
\end{IEEEbiography}

\begin{IEEEbiography}
 [{\includegraphics[width=1in,height=1.25in,clip,keepaspectratio]{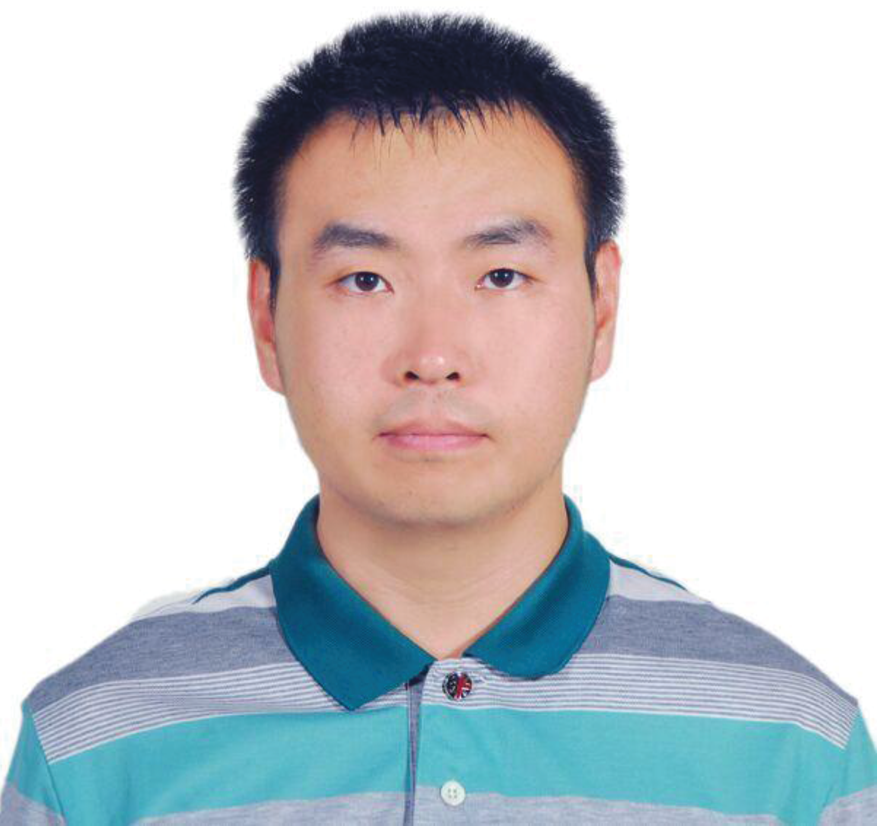}}]{Shi Yan}
 (M'19) received the Ph.D. degree in communication
and information engineering from Beijing
University of Posts and Telecommunications
(BUPT), China, in 2017. He is currently an assistant
professor in the key laboratory of universal
wireless communications (Ministry of Education)
at BUPT. In 2015, he was an Academic Visiting
Scholar with Arizona State University, Tempe, AZ,
USA. His research interests include game theory,
resource management, deep reinforcement learning,
stochastic geometry and fog radio access networks.
\end{IEEEbiography}

\end{document}